# Particle size effect on magnetic properties of interacting $La_{0.67}Sr_{0.33}MnO_3$ nanoparticles

**Ali Rostamnejadi[1, 2], Hadi salamati[1] and Parviz Kameli[1]**

[1]Department of Physics, Isfahan University of Technology, Isfahan, 84156-83111, Iran

## Abstract

Magnetic nanoparticles of $La_{0.67}Sr_{0.33}MnO_3$ (LSMO) manganite with mean particle sizes of 13, 16, 18 and 21 nm were prepared by the sol-gel method. The crystal structure and mean particle size of the synthesized powders were estimated by X-ray diffraction (XRD) analysis using rietveld refinement and transmission electron microscopy (TEM). Fourier transform infrared (FTIR) transmission spectroscopy revealed that stretching and bending modes are influenced by calcinations temperature. Dc magnetization versus magnetic field of the samples was carried out at room temperature. Magnetic dynamics of the samples was studied by the measurement of ac magnetic susceptibility versus temperature at different frequencies and ac magnetic fields. A frequency-dependent peak was observed in ac magnetic susceptibility versus temperature which is well described by Vogel-Fulcher and critical slowing down laws, and empirical $c_1 = \frac{\Delta T_f}{T_f \Delta(\log_{10} f)}$ and $c_2 = \frac{T_f - T_0}{T_f}$ parameters. By fitting the experimental data with Vogel-Fulcher and critical slowing down laws, the relaxation time, characteristic temperature, magnetic anisotropy energy, effective magnetic anisotropy constant and critical exponent zυ have been estimated. The obtained values of $c_1$, $c_2$, $T_0$ and $\tau_0$ from the Vogel-Fulcher law support of the presence of strong interaction between magnetic nanoparticles. The values of zυ and $\tau_0$ obtained from critical slowing down fit suggest the presence of superspin glass behavior in LSMO nanoparticles of different sizes.

**Keywords**: Manganite; Magnetic nanoparticles; Sol-Gel; Ac magnetic susceptibility; Superspin glass.

PACS: 75.47.Lx; 75.20.-g; 75.10.Nr; 75.75. +a; 75.50.Lk

[2] Author to whom all correspondence should be addressed. Tel.: +98 311 3912375; fax: +98 311 3912376. E-mail address: ali@ph.iut.ac.ir.

## 1. Introduction

Recently magnetic nanoparticles systems have been of great interest due to their spectacular physical properties and their technological applications such as magnetic recording media, magnetic sensors, permanent magnets, ferrofluids, magnetocaloric refrigeration, magnetic resonance imaging (MRI) enhancement, magnetically guided drug delivery and hyperthermia [1-5]. The magnetic properties of nanoparticles strongly depend on the size and shape of particles, particle size distribution, finite-size effect and dipolar or exchange interaction between the particles [6-9]. In a large number of magnetic nanoparticles applications, densely packed nanoparticles are used, thus it is important to know the effects of interaction between nanoparticles on physical properties of these systems [2, 8]. If the particle size is smaller than the size of single domain, each particle has a large magnetic moment (so-called superspin) [2, 6, 9]. The noninteracting superspins give rise to superparamagnetic behavior [6-11]. In superparamagnetic state, although the magnetic order still exists within the particles, each particle behaves like a paramagnetic atom and the magnetic nanoparticle goes through a superparamagnetic relaxation process, in which the magnetization direction of the nanoparticle rapidly fluctuates, instead of fixing along certain direction. The temperature, at which the magnetic anisotropy energy of a nanoparticle is overcome by thermal activation, is known as the blocking temperature [7-9]. When the interactions between the superspins, which are fully frustrated and random, become sufficiently strong, the system of interacting superspins shows the superspin glass behavior at below a freezing temperature [2, 6-9]. Ac magnetic susceptibility is used to study the dynamics of magnetic properties of magnetic nanoparticles. By this technique one can distinguishes between superparamagnetic and superspin glass systems [2, 6-11]. There are various phenomenological models which are used to explain the magnetic dynamics behavior of such systems based on frequency dependence of the ac magnetic susceptibility. The Néel-Brown law is applied to study the dynamics of noninteracting superspin systems. The interactions between superspins have been taken into account in the Vogel-Fulcher law, which is a modification of the Néel- Brown law [6-8, 10]. The critical slowing down law, which assumes the existence of true equilibrium phase transition with a divergence of relaxation time near the transition temperature, has been used to explain the relaxation behavior in superspin glass and spin-glass systems [8, 9, 11].

For hyperthermia application, magnetic nanoparticles of fairly uniform size, having a Curie temperature above room temperature, are needed. Manganites with a typical composition $La_{0.67}Sr_{0.33}MnO_3$ (LSMO) are of interest in this context due to its high $T_C$ value (about 380 K) and a large magnetic moment at room temperature [12-15]. In this paper we report on the preparation of nanoparticles of $La_{0.67}Sr_{0.33}MnO_3$, with different particle size, by sol-gel method. Phase formation, crystal structure and particle size were studied by X-ray diffraction (XRD) and transmission electron microscopy (TEM) respectively. The dc magnetization measurement was done at room temperature (295 K). The magnetic dynamic of magnetic LSMO nanoparticles were investigated by measuring the ac magnetic susceptibility versus temperature at different frequencies and ac magnetic fields. The phenomenological Néel- Brown, Vogel-Fulche and critical slowing down models have been used to study the dynamical properties. Results show that there are strong interaction between nanoparticles of the powder of LSMO with different sizes which assuming the presence of superspin glass behavior.

## 2. Experiment

Nanoparticles of LSMO manganite were prepared by sol-gel method. Stoichiometric amounts of the nitrate precursor reagents ($La(NO_3)_3.6H_2O$, $Mn(NO_3)_2.4H_2O$ and $Sr(NO_3)$) were dissolved in water and mixed with ethylene glycol and citric acid, forming a stable solution. The solution was then heated on a thermal plate under constant stirring at 80 ˚C to eliminate the excess water and to obtain a viscous gel [10, 16]. The gel was dried at 250 ˚C and then calcinated at 450 ˚C (S1) for 4 h. Different packages of powder were sintered at 500˚C (S2), 525˚C (S3), 550˚C (S4) and 600˚C (S5) for 4 h to obtain powders with different particle sizes. Phase formation and crystal structure of the powders were checked by XRD pattern using Cu K$\alpha$ radiation source in the 2θ scan range from 20° to 80°. The average particle sizes of the samples were estimated from the X-ray peak width by using the Scherrer's formula and TEM micrograph. The ac magnetic susceptibility has been measured versus temperature at different frequencies and ac magnetic fields in the selected range of 40-1000 Hz and 80-800 A/m respectively, using a Lake Shore ac susceptometer model 7000. The dc magnetization was carried out as a function of magnetic field at 294 K using a vibrating sample magnetometer. The infrared (IR) transmission measurement was carried out on powder samples of LSMO using a Fourier Transform IR (FTIR) JASCO 680 plus spectrophotometer. The powders of LSMO were diluted by KBr and pressed in to a disk of thickness 0.1 mm.

## 3. Results and discussion

The XRD is used to verify the crystal structure of the samples. In Fig. 1 the XRD pattern of the samples are shown. Fig. 1 shows S1 sample is amorphous and its crystal structure is not formed. By increasing the sintering temperature from 500 ˚C to 600 ˚C, the perovskite crystal structure is formed, but there is still an amorphous phase present in the S2 sample. The Rietveld analyses of the XRD pattern of the samples have been carried out using FULLPROF program [17]. Figure 2 shows the XRD pattern with Rietveld analysis of the S4 sample. The results of Rietveld analysis for S3, S4 and S5 samples are given in table 1. As can be seen from table 1, the lattice constants of the samples are increased by increasing the particle size which may be due to the slight decrease of the oxygen stoichiometry from the ideal one caused by increasing the sintering temperature. This behavior is also reported for LSMO nanoparticles [12].

The broadening of the XRD lines corresponds to the decreasing of particle size (inset Fig. 1). The average particle size, *d,* of the samples were estimated using Scherrer's formula,

$$d = \frac{k\lambda}{\beta \cos\theta} \tag{1}$$

Where k=0.9 is the particle shape factor, considering the spherical shape of the nanoparticles, λ=1.5405Å is the wavelength of Cu *Kα* radiation, β is the full width at half maximum of the XRD (104) peak, and θ is the diffraction angle of the peak [18]. The variations of particle sizes with sintering temperatures are shown in table 1. As one can see from table 1, by increasing the sintering temperatures the particle sizes are increased. The TEM micrograph of S3 and S5 samples are shown in Fig. 3. TEM micrographs show that the particle size distribution is almost homogenous and the mean particle sizes of the S3 and S5 samples are about 16 and 23 nm respectively and they are in agreement with the mean particle sizes of these samples estimated from the Scherrer's formula (table 1).

Figure 4 shows the FTIR transmission spectra of the diluted powders of the S1, S2, S3, S4 and S5 samples at room temperature. In Fig. 4 the absorption peaks around 1650 $cm^{-1}$ and 2350 $cm^{-1}$ are of the carrier KBr $(H_2O)_n$ and $CO_2$, respectively[19]. In S1, S2 and S3 samples, the two strong peaks at 860 $cm^{-1}$ and 1450 $cm^{-1}$ belong to $SrCO_3$ [20, 21]. Calcinations of the powders at 550 ˚C and 600 ˚C will decrease the above peaks; this is the indication of decomposition of $SrCO_3$ at 550 ˚C. The absorption peaks around $\nu_3 \approx 600\, cm^{-1}$ and $\nu_4 \approx 400\ cm^{-1}$ should belong to stretching, $\nu_3$ and bending, $\nu_4$ of the internal phonon modes of $MnO_6$ octahedral [19-21]. The stretching mode is related to the change of Mn-O-Mn bond length and the bending mode involves the change of Mn-O-Mn bond angle. The appearance of the stretching and bending modes at transmission spectra indicates that the perovskite structure of LSMO has been formed at temperature of 500 ˚C, which is in agreement with the result of XRD.

To study the magnetic behavior of the samples, the magnetic field dependence of magnetization has been measured at room temperature (295 K). The field dependence of the magnetization for S2, S3, S4, and S5 samples are shown in Fig. 5. From Fig. 5, the M-H curve of S2 sample shows superparamagnetic behavior, without noticeable remanence and coercivity. However the M-H curve of S3, S4, and S5 samples show that the magnetization has a ferromagnetic type behavior with a small hysteresis loop and a low coercive field. The fact that magnetization is not saturated in the magnetic fields up to 8.5 kOe, shows that some of magnetic nanoparticles of S3, S4, and S5 sample are in the superparamagnetic state while the existence of the coercive field indicates that the rest of particles are blocked due to the overcoming of the thermal energy by their anisotropy energy. This is due to the existence of size distribution of magnetic nanoparticles and consequently these samples have different blocking/freezing temperatures. As will be seen, this is in agreement with the results of ac magnetic susceptibility measurement. We have also applied the well-known Arrott-Belov-Kouvel (ABK) plot ($M^2$ versus H/M) for S2, S3, S4 and S5 samples as shown in the lower insets of Fig. 5. From Fig. 5, ABK plot shows a strong convex curvature with a finite spontaneous magnetization which is a signature of ferromagnetic phase of the samples [22, 23]. Spontaneous magnetization can be estimated by linear fitting of the high magnetic field part of the $M^2$-H/M curve and intercepts it with the $M^2$ axis. By increasing the particle size, the spontaneous magnetization and curvature are increased which indicates that the magnetic order in the samples are increased. This behavior has been reported previously for LSMO nanoparticles [22, 23].

Ac magnetic susceptibility measurement is one of the standard methods which are used to obtain the information on the dynamical properties of magnetic nanoparticles [6-10]. By this technique the effect of ac and dc magnetic field and frequency on blocking/freezing temperature can be investigated. The temperature dependence of the ac magnetic susceptibility and zero field cooled (ZFC) magnetization of magnetic nanoparticles show a characteristic maximum which is the signature of blocking/freezing process of the superparamagnetic/spin glass systems [6-11]. This peak is also observed in nanoparticles of manganites and its nature is described in different forms. The peak in the magnetization measurement of $La_{2/3}Ca_{1/3}MnO_3$ nanoparticles versus temperature by Li et al. [24] have been referred as a superparamagnetic behavior but Markovic et al. [25] have reported the same peak as a spin glass behavior. In the nanoparticles of $Nd_{0.7}Sr_{0.3}MnO_3$ [26] and $La_{0.6}Pb_{0.4}MnO_3$ [27], the behaviors of the systems have been referred as cluster glass like and superparamagnetic respectively. For nanoparticles of $La_{2/3}Sr_{1/3}MnO_3$ [28] and $La_{0.8}Sr_{0.2}MnO_3$ [29] superparamagnetic properties have been reported. To understand the nature of this peak in nanoparticles of LSMO manganite with different sizes, we have used the ac magnetic susceptibility measurement at different frequencies and ac magnetic fields.

Figure 6 shows the ac magnetic susceptibility of S2, S3, S4 and S5 samples versus temperature at an ac magnetic field of 800 A/m and frequency of 1000 Hz. As can be seen from Fig. 6, ac susceptibility measurements show a characteristic peaks between 263and 315 K, which is due to the blocking/freezing of nanoparticles of LSMO with different sizes. As indicated in the inset of Fig. 6(a) the position of this peak is increased by increasing the particle size. Another peak is also observed in the ac magnetic susceptibility in the temperature range between 180-200 K for S2, S3, S4 and S5 samples. The nature of this peak can be explained in terms of the core-shell model [25]. In this model, each particle can be supposed to consist of two different parts, an inner core and a shell. The magnetic properties of the cores are the same as the bulk sample. On the other hand the magnetic properties of the shells are different from the bulk and depend on particle size and surface effects such as vacancies, stress, defects and broken chemical bonds. In the case of our samples, the nominal valance of the Mn ions in the core is the same as the stoichiometric bulk counterpart. Consequently, the double exchange interaction between $e_g$ electrons causes ferromagnetic behavior in the core. In the shell, surface effects modify the magnetic interactions by localization of $e_e$ electrons and causes different valances for Mn ions. Therefore the ferromagnetic (double exchange) interaction and antiferromagnetic (superexchange) interaction exist simultaneously in the shell. Because of the existence of disorder in the interaction and position of the ions in the shell, a surface spin freezing state can be introduce below a freezing temperature. Therefore the upper peak is due to the blocking/freezing of the core spins and the lower peak is due to the surface spin freezing [30, 31]. Recently the surface spin freezing for nanoparticles of LSMO has been reported [28].

To study the magnetic dynamic behavior of the S2, S3, S4 and S5 samples, we measure the ac magnetic susceptibility versus temperature at different frequencies. Figure 7 shows the temperature dependence of the real, $\chi'(T)$ and imaginary, $\chi''(T)$ parts of ac magnetic susceptibility of the S5 sample at different frequencies in the range of 40-1000 Hz and at an ac magnetic field of 800 A/m. In Fig. 7, $\chi'(T)$ and $\chi''(T)$ show a peak near room temperature which is frequency dependent and shifted to higher temperature with increasing frequency. The frequency dependence of the ac magnetic susceptibility is a characteristic of superparamagnetic/spin glass systems [2, 6-11].

There are three well known phenomenological models which are used to describe the dynamical behavior of magnetic nanoparticles and distinguish between superparamagnetic and superspin glass systems. For noninteracting nanoparticles the frequency dependence of blocking temperature has been given by Néel-Brown model [6],

$$\tau = \tau_0 \exp(\frac{E_a}{k_B T}) \qquad (2)$$

where $\tau$ is related to measuring frequency ($\tau$=1/f) and $\tau_0$ is related to the jump attempt frequency of the magnetic moment of nanoparticle between the opposite directions of the magnetization easy axis. For superparamagnetic systems $\tau_0$ is in the range of $10^{-9}$-$10^{-13}$ s [3, 6-8]. In the absence of external magnetic field and interaction between nanoparticles, the energy of barrier, $E_a$ can be assumed to be proportional to particle volume V and can be written as [3]

$$E_a=k_{eff}V\sin^2\theta \quad (3)$$

here $k_{eff}$ is an effective magnetic anisotropy constant and θ is the angle between magnetic moment of particle and its easy axis. Blocking temperature is the one that the thermal energy overcomes to anisotropy energy. When the energy of the potential barrier is comparable to thermal energy, the magnetization direction of the nanoparticles starts to fluctuate and goes through a rapid superparamagnetic relaxation. Above $T_B$ the magnetization direction of nanoparticles can follow the direction of the applied field. Below the blocking temperature the thermal energy is less than the anisotropy energy, hence the direction of magnetization of each nanoparticle which may lie in the direction of easy axis, is blocked. Since the nanoparticles and consequently their easy axes are randomly oriented, by decreasing the temperature the total magnetic susceptibility is reduced. By fitting the experimental data from ac magnetic susceptibility of S2, S3, S4 and S5 samples with Eq. (2), (Fig. 8), we have found an unphysical low values for $\tau_0$ (table 2) in comparison to the values of $10^{-9}$ -$10^{-13}$ for superparamagnetic systems. As expected, this result simply indicates that there exists strong interaction between nanoparticles of S2, S3, S4 and S5 samples.

For noninteracting magnetic nanoparticles, blocking temperature can be estimated by relation [4]

$$T_B= kV/25k_B \quad (4)$$

where V is the average volume of nanoparticles, $k_B$ is the Boltzmann constant and k is the magnetic anisotropy constant which for single crystal of LSMO is $1.8\times10^4$ erg/cm$^3$ [32]. Using the average particle size from table 1, one obtains the value of $T_B$ about 6 K, 11 K, 16 K and 25 K for S2, S3, S4 and S5 samples, respectively. These values for blocking temperatures are much lower than the values that we have observed in Fig. 6 for S2, S3, S4 and S5 samples. These results also show that there exists strong interaction between LSMO nanoparticles with different sizes.

The interaction between nanoparticles affected the blocking/freezing temperature by modifying the potential barrier [2, 6, 7]. By increasing the strength of interaction, $T_B$ shifts to higher temperatures.

For interacting magnetic nanoparticles, the frequency dependence of $T_B$ is given by the Vogel-Fulcher law [6],

$$\tau = \tau_0 \exp\left(\frac{E_a}{k(T - T_0)}\right) \quad (5)$$

here $T_0$ is an effective temperature which reveals the existence of the interaction between nanoparticles and T is the characteristic temperature indicating the onset of the blocking process (i.e. the temperature of peak position in the imaginary component of ac magnetic susceptibility). In Fig. 9 we tried to fit the experimental data of $\chi''(T)$ for our samples, using Eq. (5). The obtained results from this fitting for $T_0$, $\tau_0$ and $E_a/k$ are given in table 2. A good agreement of experimental data with the Vogel-Fulcher law is the evidence that the phenomenon that occurs at $T_B$ is related to blocking/freezing of an ensemble of interacting nanoparticles. Further, from the fitting of experimental data with Eq. (5) and using the average particle sizes from table 1, one can obtain the value of magnetic anisotropy constants for S2, S3, S4 and S5 samples. These values, which are shown in Fig. 10, are slightly larger than the value of magnetic anisotropy constant of a single crystal of LSMO manganite [32]. This difference may be due to the existence of other sources of magnetic anisotropies like shape anisotropy and surface anisotropy or magnetic interaction between nanoparticles. By increasing the particle size the magnetic anisotropy constant is reduced. This result suggests that surface effects on magnetic properties of LSMO nanoparticles are very important.

To classify the observed blocking/freezing process, two useful parameters $c_1$ and $c_2$ are usually used [6, 7],

$$c_1 = \frac{\Delta T_f}{T_f \Delta(\log_{10} f)} \tag{6}$$

$$c_2 = \frac{T_f - T_0}{T_f} \tag{7}$$

here $\Delta T_f$ is the difference between $T_f$ measured at the frequency $\Delta(\log_{10} f)$ interval, $T_f$ is the mean value of blocking temperature in the range of experimental frequencies and $T_0$ is the characteristic temperature of the Vogel-Fulcher law, Eq. (5). The value of $c_1$, which is independent of any model, represents the relative shift of blocking temperature per decade of frequency. The value of $c_2$ can be useful to compare the $T_f$ variation between various systems. The experimentally values of $c_1$ and $c_2$ depend on the interaction strength between magnetic nanoparticles. Dormann et al. distinguish three different types of dynamical behavior based on the values of $c_1$ and $c_2$: (1) for noninteracting particles $0.1 < c_1 < 0.13$ and $c_2=1$ (theory), (2) for weak interaction regime (inhomogeneous freezing) $0.03 < c_1 < 0.06$ and $0.3 < c_2 < 0.6$ and in the medium to strong interaction regime (homogeneous freezing) $0.005 < c_1 < 0.02$ and $0.07 < c_2 < 0.3$ [6, 7]. Both $c_1$ and $c_2$ decrease with increasing the interactions between nanoparticles. The calculated values of $c_1$ and $c_2$ for our samples are given in table 2. By comparing the values of $c_1$ and $c_2$ for the S2, S3, S4, and S5 samples in table 2 with the above values, one can claim the presence of superspin glass behavior in nanoparticles of LSMO with different sizes.

We have checked the possibility of superspin glass behavior based on ac magnetic susceptibility by conventional critical slowing down model [2, 8]. In this model the characteristic relaxation time τ diverges at the transition temperature according to

$$\tau = \tau_0 (T / T_g - 1)^{-z\nu} \qquad (8)$$

where $T_g$ is the transition temperature, $\tau_0$ is related to the relaxation time of the individual particle magnetic moment, ν is the critical exponent of correlation length, $\xi \approx (T / T_g - 1)^{-\nu}$ and z relates τ and ζ as $\tau \propto \xi^z$ [8]. The divergence of the correlation length or equally, relaxation time near $T_g$ indicates the presence of a true equilibrium thermodynamic phase transition. The log-log plot of the external frequency (f) versus reduced temperature, $(T-T_g)/T_g$, for S2, S3, S4 and S5 samples which are shown in Fig. 11, gives an excellent linear dependence. The obtained values of $\tau_0$, $T_g$ and zν are given in table 2. The typical values of $\tau_0$ and zν for spin-glass systems are in the range of $10^{-9}$-$10^{-13}$ s and 7-12, respectively [33, 34], while for interacting nanoparticle systems the smaller values are also reported. For nanoparticles of $Nd_{0.7}Ba_{0.3}MnO_3$ with particle sizes of 20 nm and 41nm, the values of $\tau_0 \approx 10^{-6}$ and zν=6.03, and $\tau_0 \approx 10^{-5}$ and zν=5.3 has been reported, respectively [35]. For interacting nanoparticles of $\gamma$-$Fe_2O_3$ the values of $\tau_0 \approx 10^{-9}$ and zν=10 [11], $\tau_0 \approx 10^{-11}$ and zν=7.6 [36] and in the case of spin cluster in amorphous $Fe_2O_3$ the values of $\tau_0 \approx 10^{-11}$ and zν=5.3 have been reported [37]. For NiO core-shell nanoparticles, the values of $\tau_0 \approx 10^{-12}$ and zν=8 [38] and for $Co_{50}Ni_{50}$ nanoparticles embedded in the amorphous $SiO_2$ host, the values of $\tau_0 \approx 10^{-12}$ and zν=8 [39] and for $Fe_3O_4$ nanoparticles $\tau_0 \approx 10^{-9}$ and zν=8.2 [9] have also been reported. Therefore the obtained values of zυ and $\tau_0$ of our samples are consistent with those expected for spin glass systems and suggest the existence of a phase transition toward a superspin glass state below peak temperature.

Figure 12 shows $\chi''(T)$ data of S2 sample as a function of temperature in different amplitude of ac magnetic fields in the range of 80-800 A/m and at frequency of 111.1 Hz. From Fig. 12, the ac magnetic susceptibility is strongly depends on ac magnetic field amplitude. By increasing the amplitude of ac magnetic field, the blocking temperature shifts to lower temperatures and the magnitude of susceptibility increases. These features are also signatures of the superparamagnetic/spin glass systems [31]. The applied magnetic field reduces the height of potential barrier; therefore less thermal energy is needed to overcome the anisotropy energy and consequently the blocking/freezing temperature is reduced.

## 4. Summery and conclusions

We now summarize the magnetic results to reach a conclusion about the magnetic behavior of LSMO nanoparticles.

The M-H curves show that there exist the ferromagnetic behavior in the nanoparticles of the samples and the ABK plots indicate that by increasing the particle size the ferromagnetic order is increased. The temperature dependence of ac magnetic susceptibility of the samples shows a characteristic peak which is frequency and ac magnetic field dependence. These behaviors are characterizes of superparamagnetic/spin glass systems. Although it is difficult to distinguish between an interacting superparamagnetic and a real spin glass systems, but based upon the interparticle interaction strength, there are some empirical criterions (c1 and $c_2$ parameters) and phenomenological models (Neel-Brown, Vogel-Fulcher and critical slowing down) which are used to characterize the magnetic dynamic behavior of magnetic nanoparticle systems. The Néel-Brown and the Vogel-Fulcher laws are used for superparamagnetic systems. The Vogel-Fulcher law is also used for superspin glasses and spin glasses as well [37, 40-42]. Fitting the experimental data with Neel-Brown model gives unphysical low values for relaxation time ($\tau_0 \approx 10^{-54}$ -$10^{-128}$ s) of the samples and indicates that there are strong interactions between nanoparticles of LSMO. The good agreement between the experimental data with Vogel-Fulcher model confirms the existence of strong interaction between nanoparticles of LSMO. By fitting the experimental data with this model, the relaxation time and magnetic anisotropy constant have been obtained for our samples with different particle sizes. The obtained results are in agreement with the reported results. The obtained values of $c_1$ and $c_2$ also reveal the presence of strong interaction between nanoparticles and support the existence of superspin glass behavior in our samples. The experimental data are well fitted by critical slowing down model which indicates a true thermodynamic phase transition in the sample by reducing the temperature. Therefore the analysis of the ac magnetic susceptibility of the samples with phenomenological models show the existence of strong interaction between nanoparticles of our samples and suggest the presence of superspin glass behavior in these samples.

**Figure captions**

Fig. 1. The XRD pattern of S1, S2, S3, S4 and S5 samples at room temperature.

Fig. 2. The observed and calculated (Rietveld analysis) XRD patterns of S4 sample at room temperature.

Fig. 3. The TEM micrographs of the S3 (a) and S5(b) samples.

Fig. 4. FTIR transmission spectra of the diluted powders of the S1, S2, S3, S4 and S5 samples at room temperature.

Fig. 5. Dc magnetization versus applied magnetic field at 295 K. (a) S2 sample, (b) S3 sample, (c) S4 sample and (d) S5 sample. The upper insets show a low hysteresis loop with a small coercive field and the lower insets show the ABK plot of the samples.

Fig. 6. Ac magnetic susceptibility of S2, S3, S4 and S5. (a) Real part and (b) Imaginary part. Inset (b): blocking temperature versus particle size.

Fig. 7. Ac magnetic susceptibility of S5 versus temperature at different frequencies.(a) Real part and (b) Imaginary part.

Fig. 8. ln(f) versus 1/T for S2, S3, S4 and S5.

Fig. 9. ln(f) versus 1/ $(T-T_0)$ for S2, S3, S4 and S5.

Fig. 10. Effective magnetic anisotropy constant of S2, S3, S4 and S5.

Fig. 11. log-log plot of the external frequency (f) versus reduced temperature, $(T-T_g)/T_g$, for S2, S3, S4 and S5 samples.

Fig 12. Temperature dependence of the imaginary part of ac magnetic susceptibility of S2 sample at different ac magnetic fields. Inset shows the blocking temperature versus applied magnetic field.

## Table captions

Table 1. The results of Rietveld refinement of XRD patterns and mean particle size of S2, S3, S4 and S5 samples.

Table 2. Physical parameters of S2, S3, S4 and S5 samples obtained from the relations 2, 5-8.

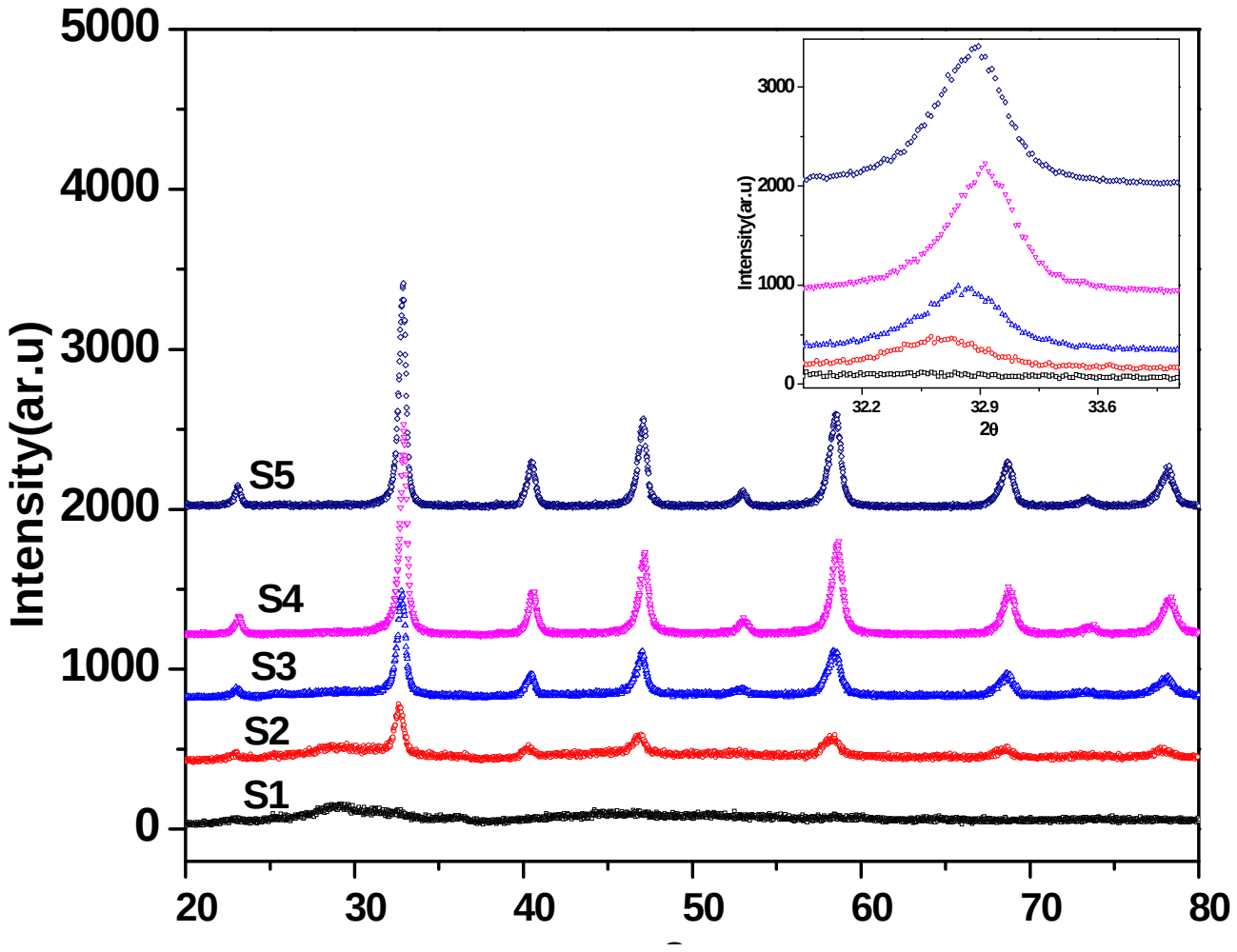

Fig. 1

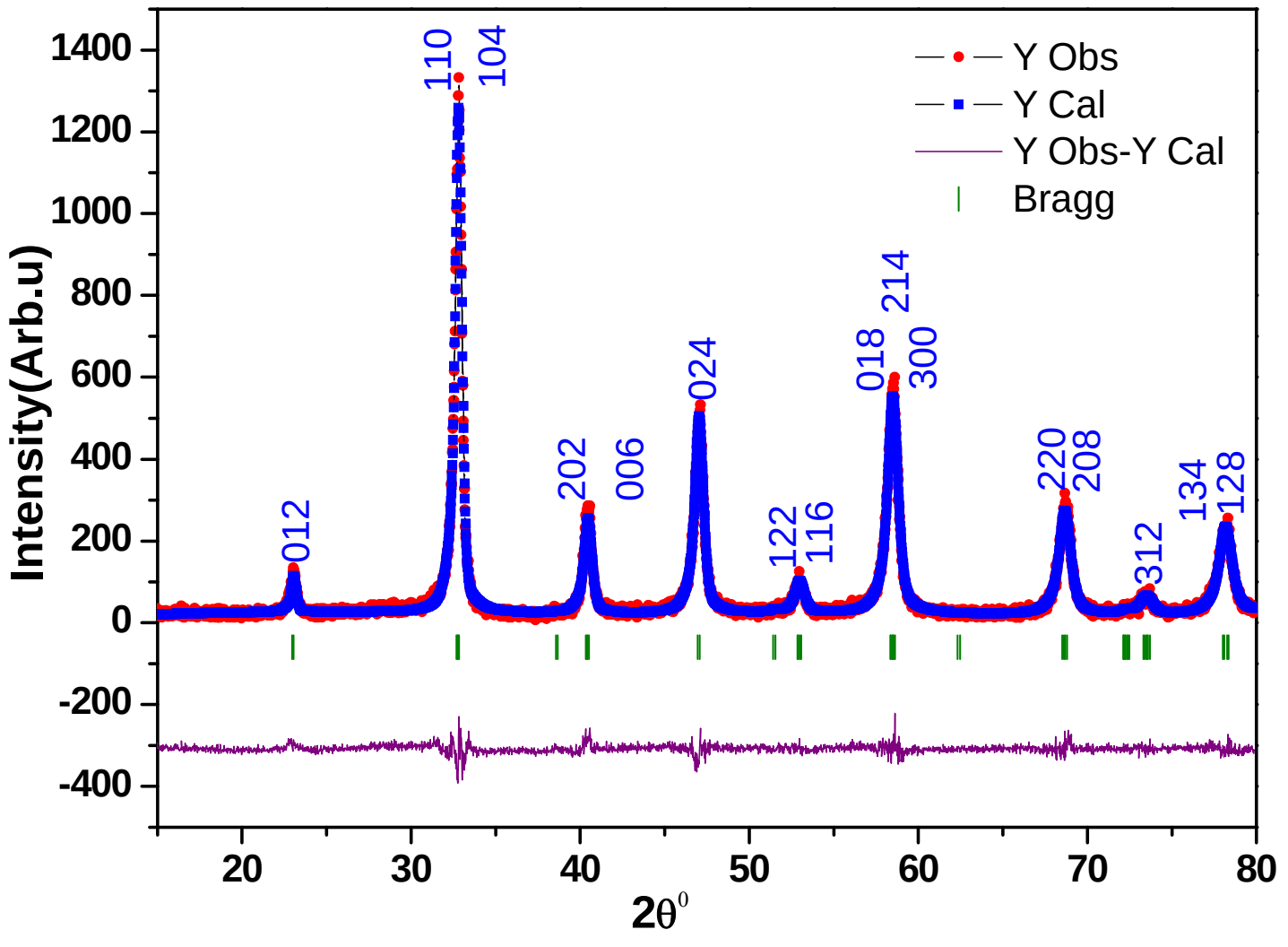

Fig. 2

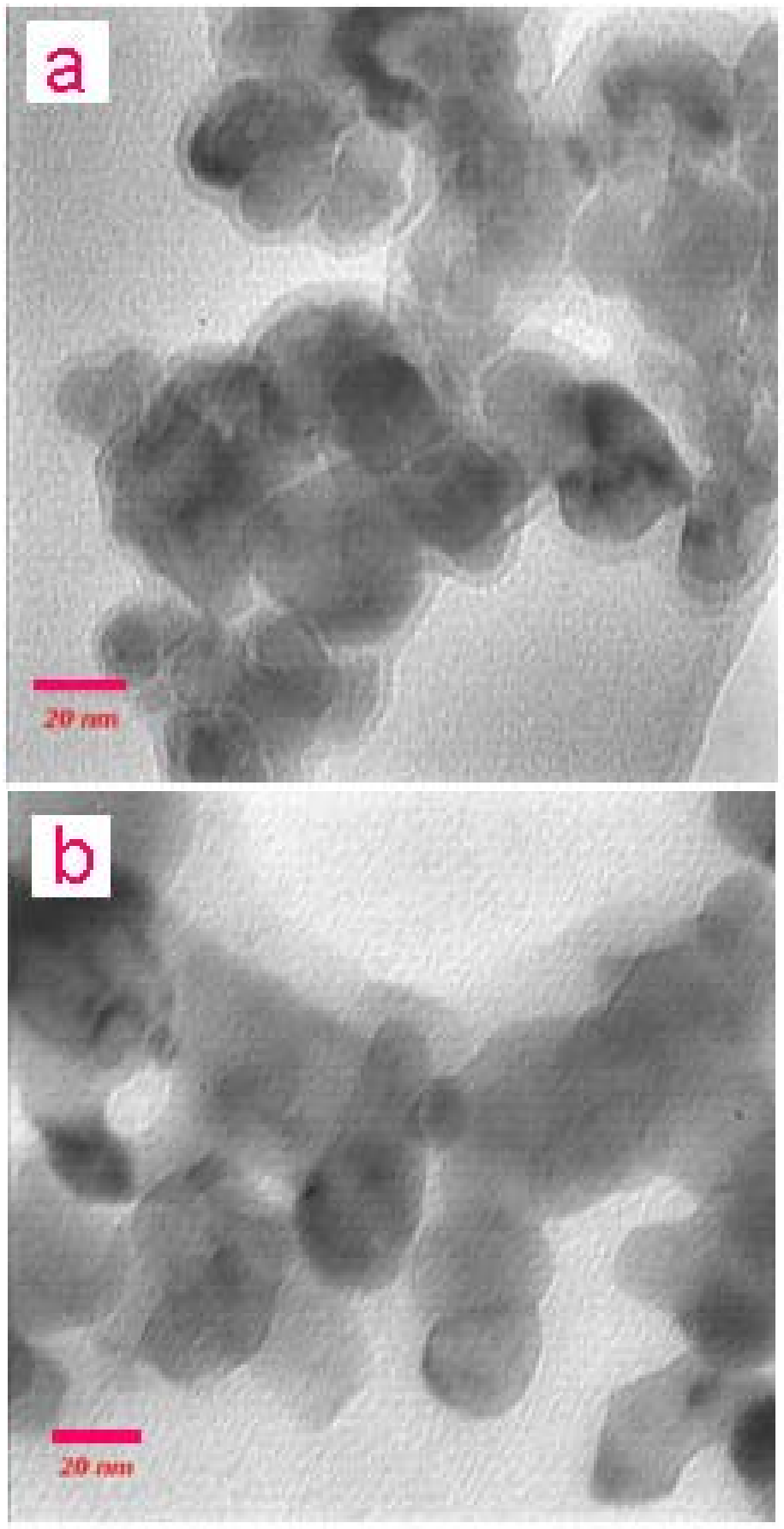

Fig. 3

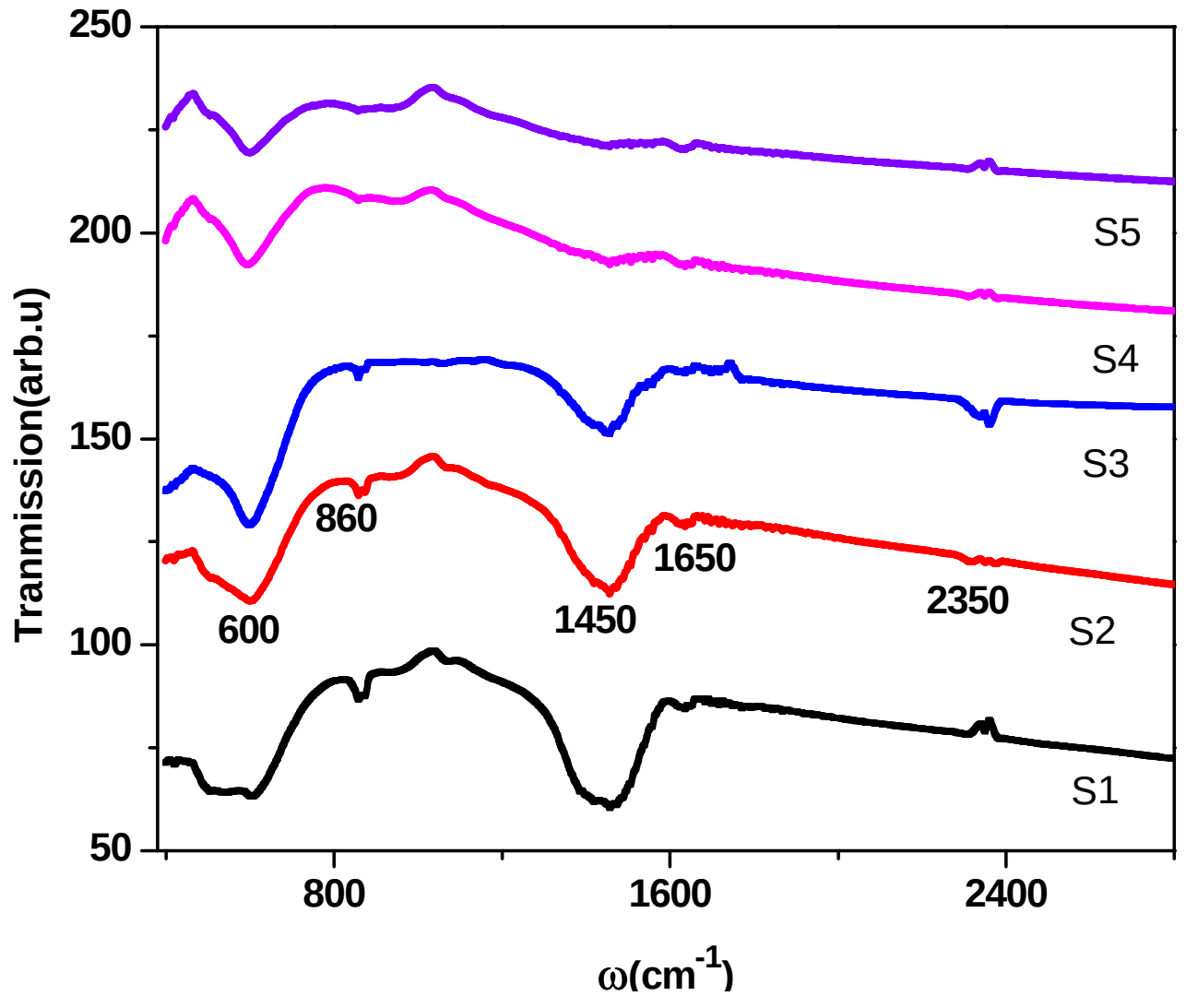

Fig. 4

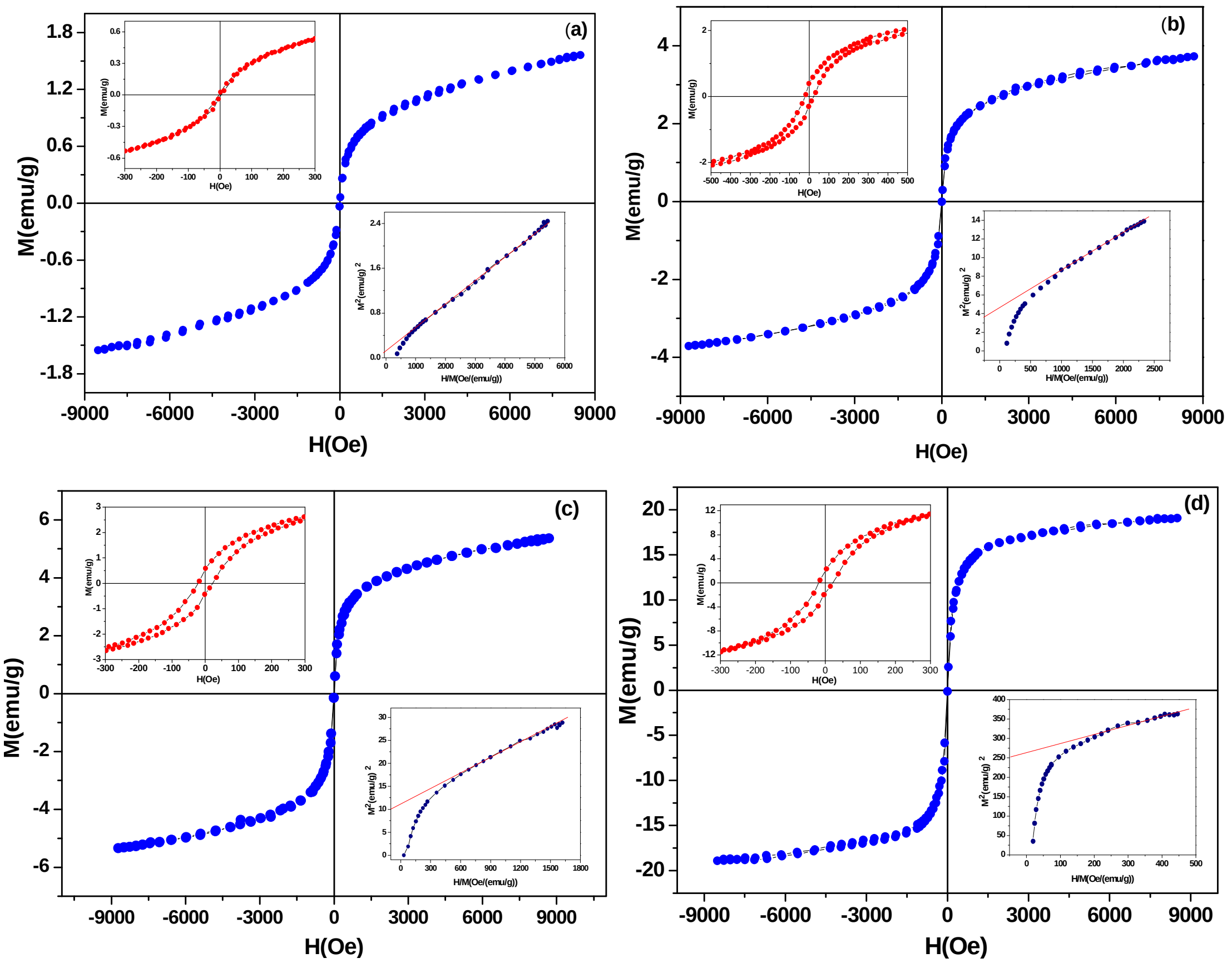

Fig. 5

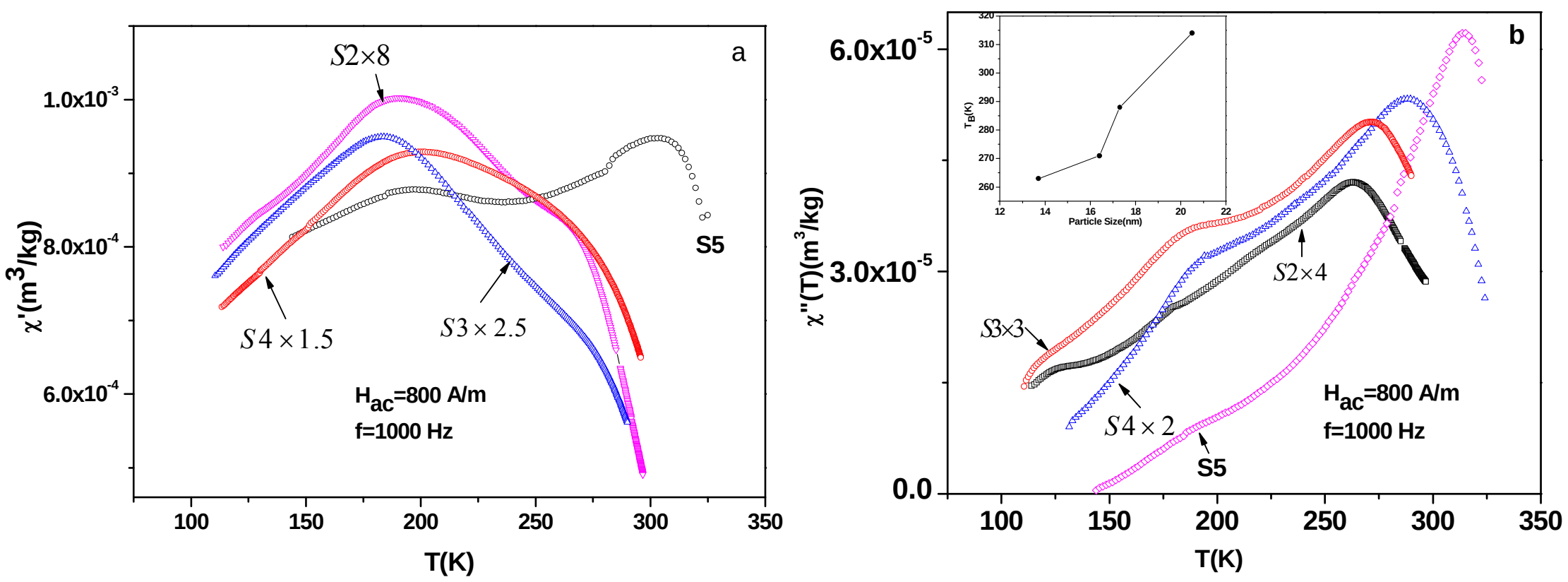

Fig. 6

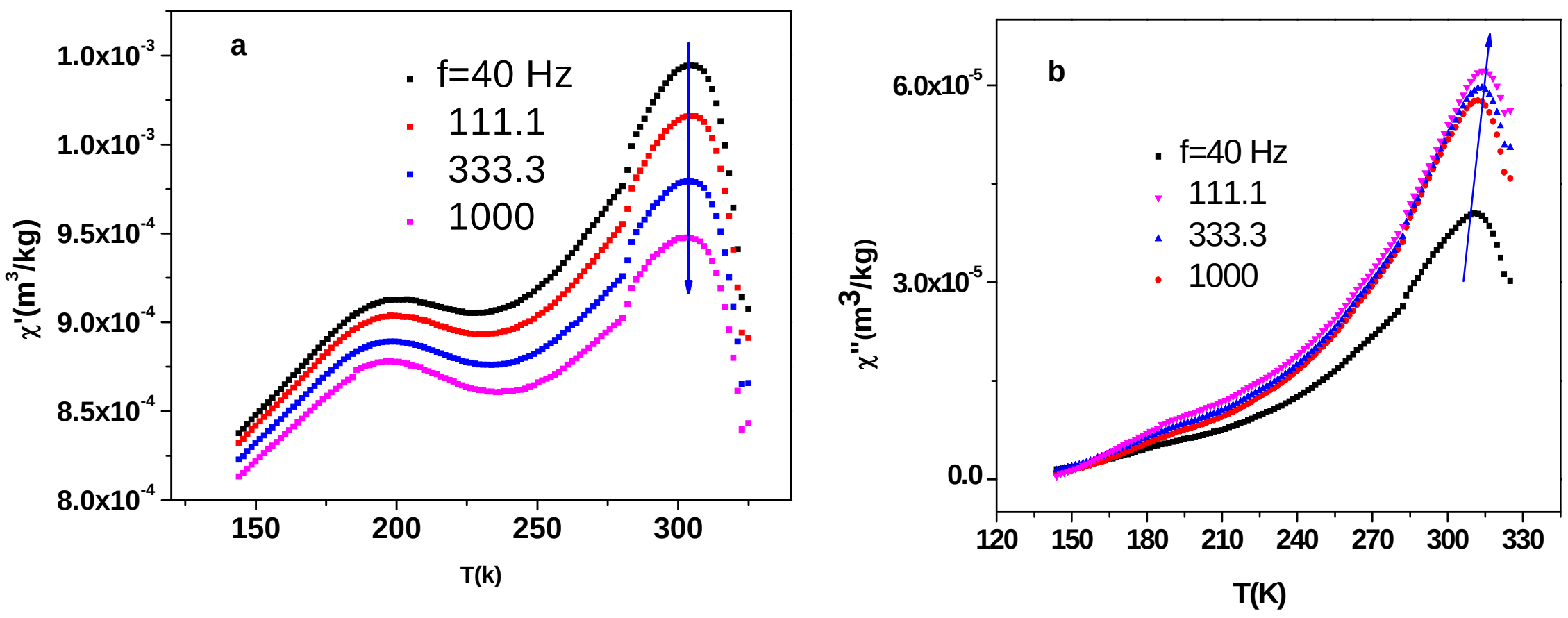

Fig. 7

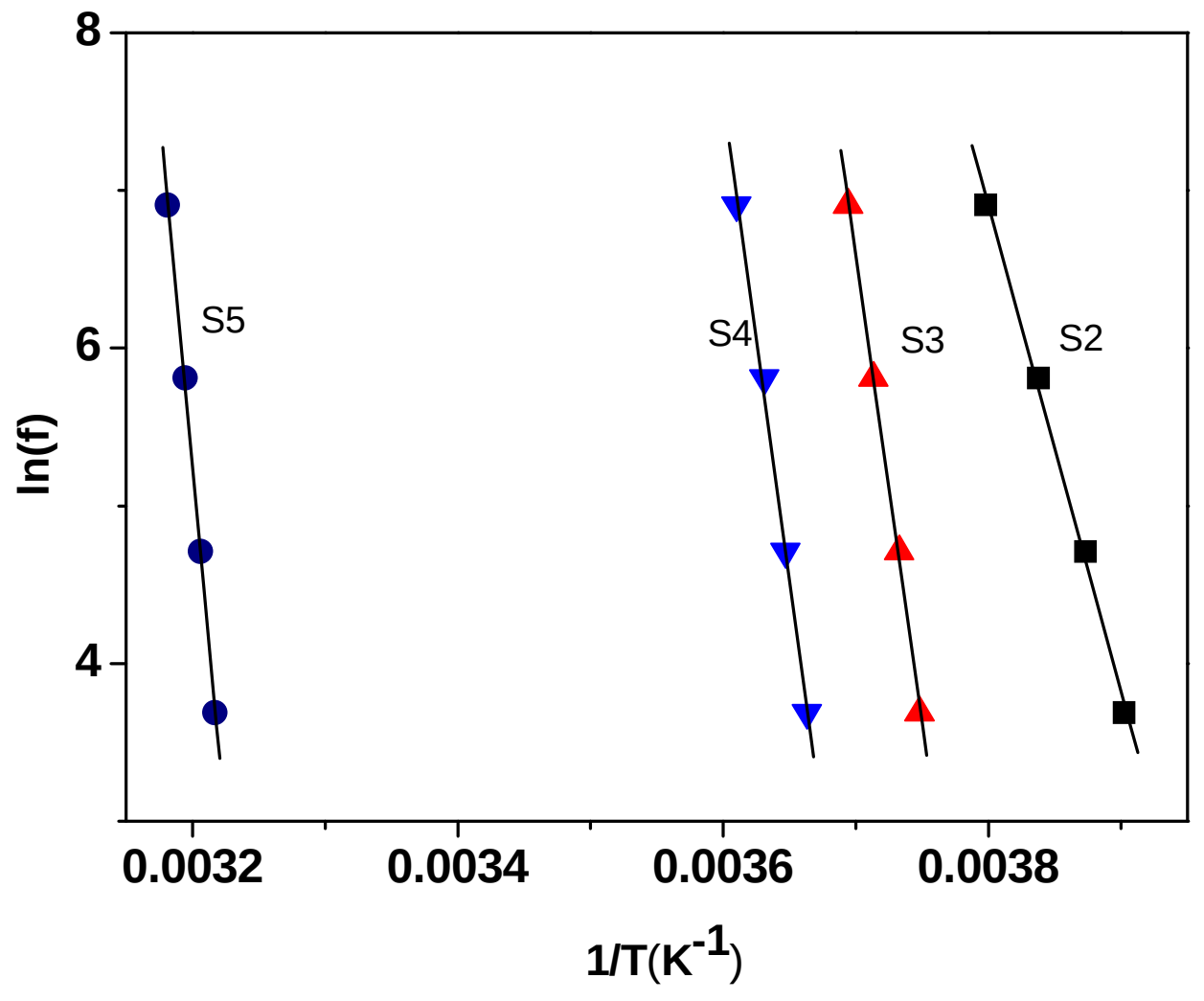

Fig. 8

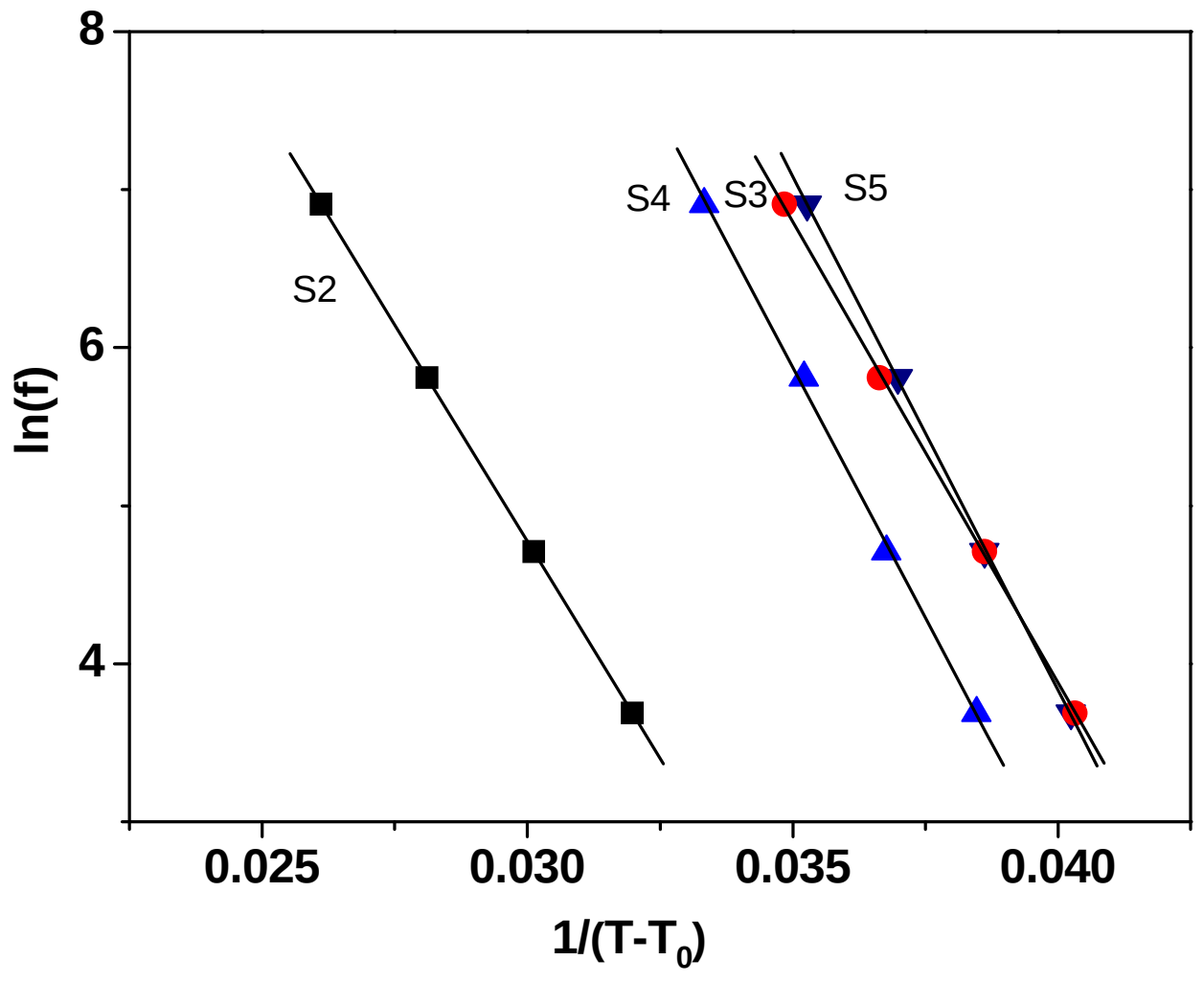

Fig. 9

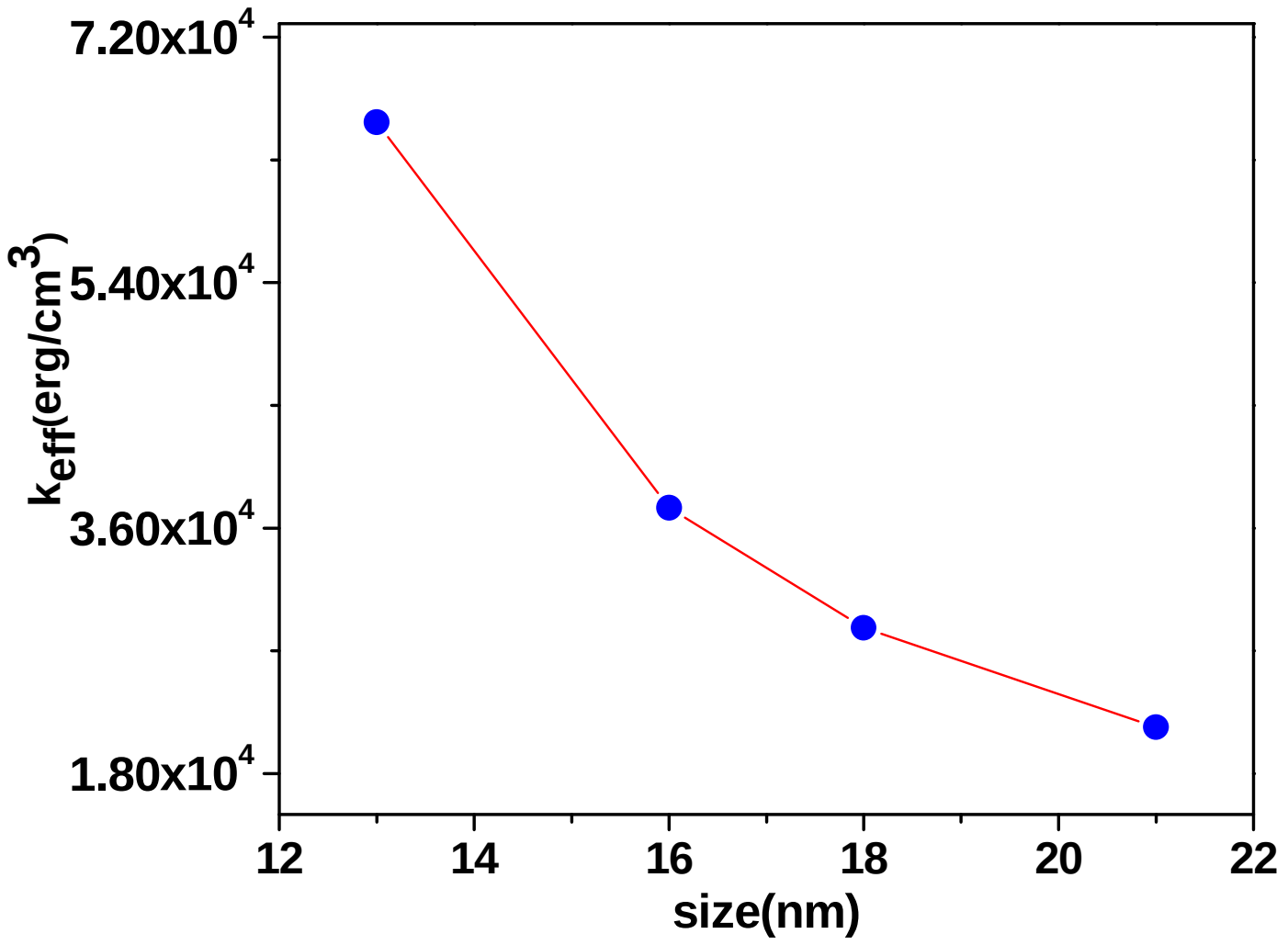

Fig. 10

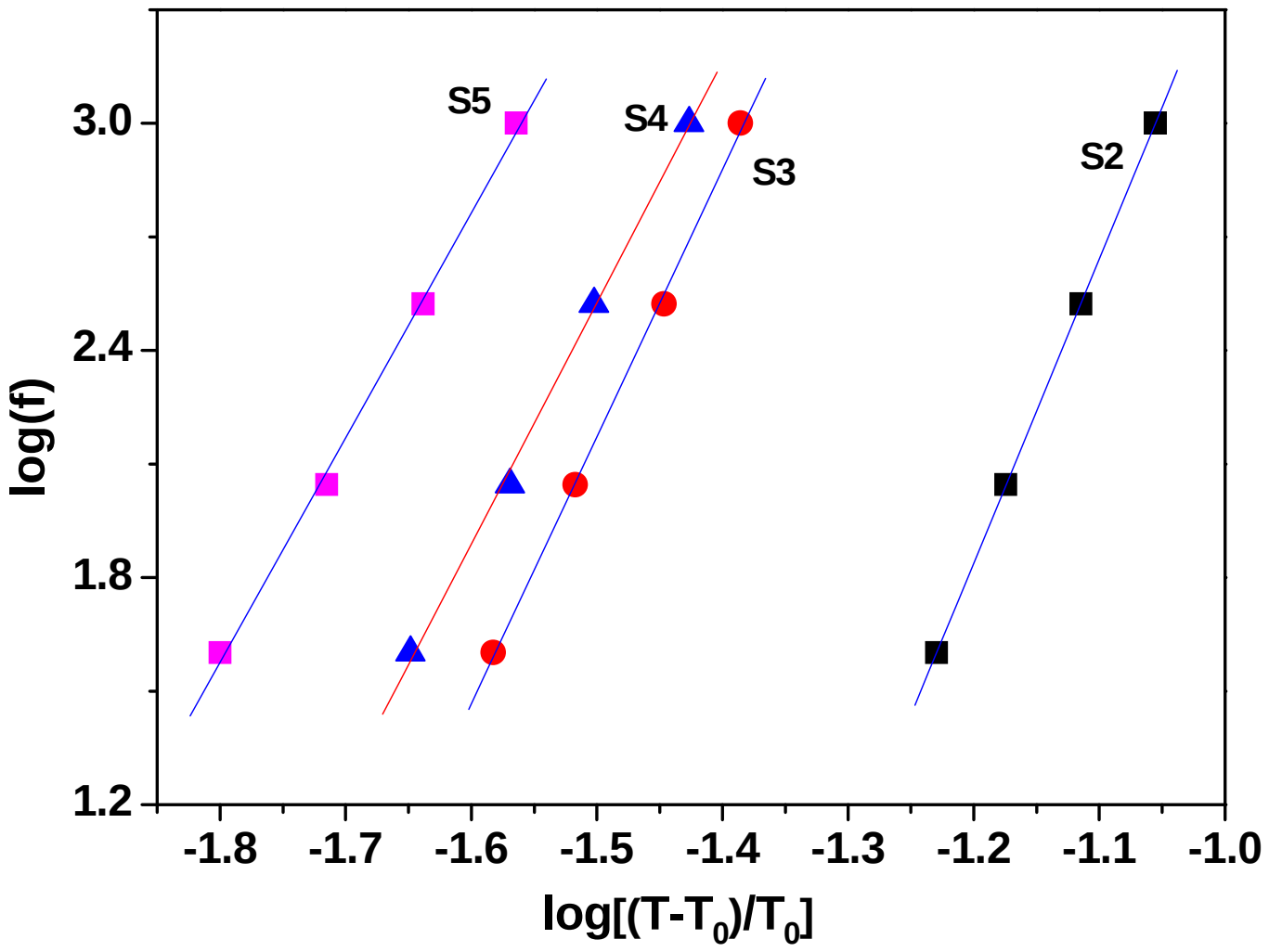

Fig. 11

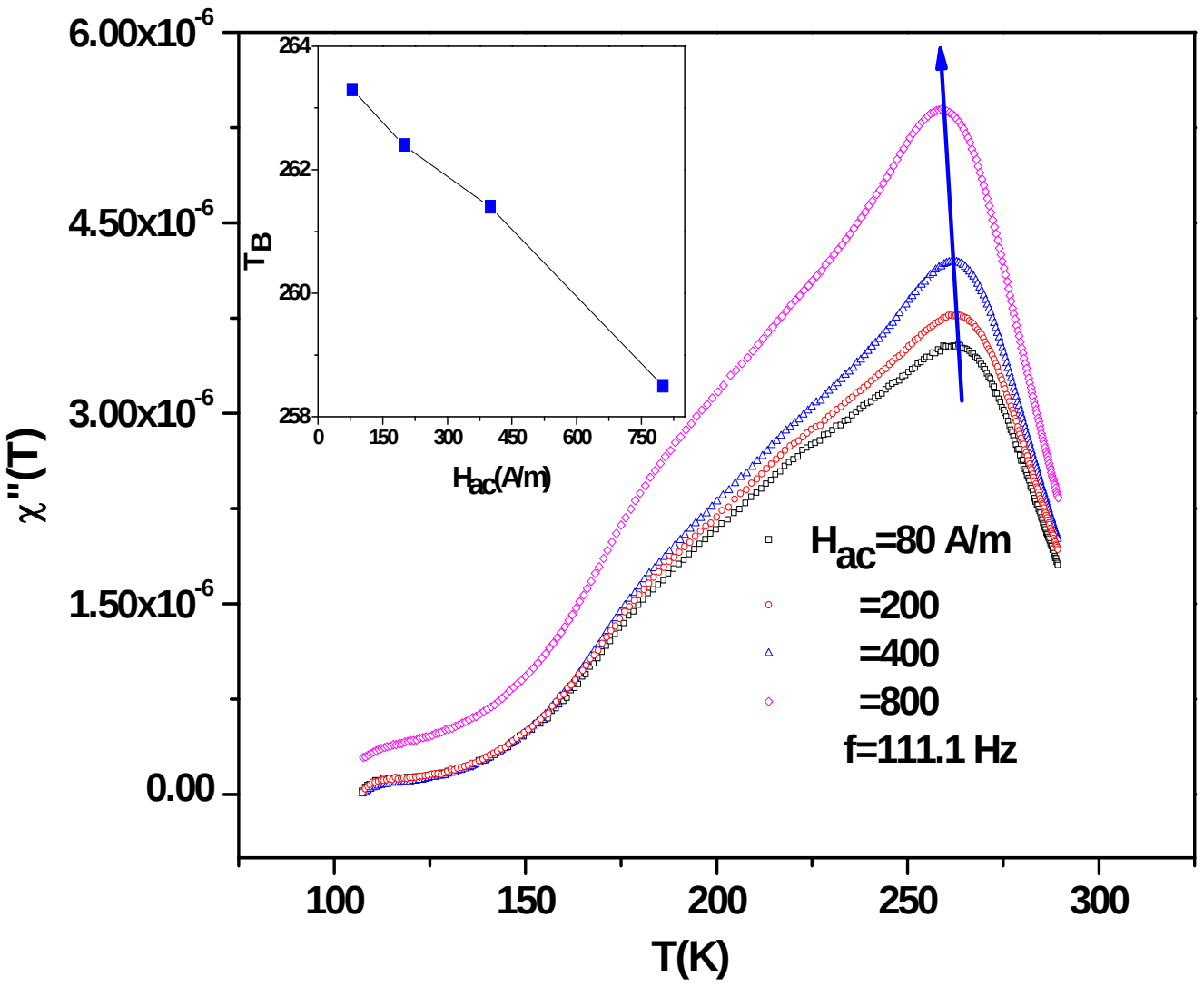

Fig 12.

Table 1

| | S2 | S3 | S4 | S5 |
|---|---|---|---|---|
| a(Å) | - | 5.4737 | 5.4752 | 5.4802 |
| b(Å) | - | 5.4737 | 5.4752 | 5.4802 |
| c(Å) | - | 13.3715 | 13.3879 | 13.4091 |
| Volume$(Å)^3$ | - | 346.9544 | 347.5653 | 348.9853 |
| Space group | - | R-3C | R-3C | R-3C |
| Mean particle size (nm) | 13 | 16 | 18 | 21 |

Table2

| model | parameter | S2 | S3 | S4 | S5 |
|---|---|---|---|---|---|
| Model-independent | $c_1$ | 0.019 | 0.01 | 0.01 | 0.008 |
| Related to Vogel-Fulcher | $c_2$ | 0.133 | 0.099 | 0.10 | 0.084 |
| Neel-Brown | $\tau_0$(s) | $1.\ 5\times10^{-54}$ | $1.1\times10^{-98}$ | $5.9\times10^{-100}$ | $2.8\times10^{-128}$ |
| | $E_a/k_B$(K) | 30806 | 59180 | 61356 | 90145 |
| Vogel-Fulcher | $\tau_0$(s) | $6.1\times10^{-10}$ | $1.5\times10^{-12}$ | $6.4\times10^{-13}$ | $1.1\times10^{-13}$ |
| | $E_a/k_B$(K) | 548 | 584 | 634 | 651 |
| | $T_0$(K) | 225 | 242 | 247 | 286 |
| Critical slowing down | $\tau_0$(s) | $3.3\times10^{-12}$ | $1.75\times10^{-13}$ | $8.4\ \times10^{-13}$ | $5.3\times10^{-13}$ |
| | $z\upsilon$ | 8.034 | 7.06 | 6.36 | 5.94 |
| | $T_g$(K) | 242 | 252 | 263 | 308 |